\documentclass{sig-alternate}

\pdfpagewidth=\paperwidth
\pdfpageheight=\paperheight

\usepackage{graphicx}
\usepackage{subfigure}

\DeclareMathOperator*{\argmin}{arg\,min}
\DeclareMathOperator*{\Tr}{Tr}
\DeclareMathOperator*{\BigO}{O}
\DeclareMathOperator*{\sgn}{sgn}

\begin{document}
%
\conferenceinfo{SIGIR'10,} {July 19--23, 2010, Geneva, Switzerland.} 
\CopyrightYear{2010}
\crdata{978-1-60558-896-4/10/07}
\clubpenalty=10000
\widowpenalty = 10000 

\title{Self-Taught Hashing for Fast Similarity Search}

\numberofauthors{4}
\author{
	\alignauthor Dell Zhang\\
		\affaddr{DCSIS}\\
		\affaddr{Birkbeck, University of London}\\
		\affaddr{Malet Street}\\
		\affaddr{London WC1E 7HX, UK}\\
		\email{dell.z@ieee.org}
	\alignauthor Jun Wang\\
		\affaddr{Dept of Computer Science}\\
		\affaddr{University College London}\\
		\affaddr{Gower Street}\\
		\affaddr{London WC1E 6BT, UK}\\
		\email{jun.wang@cs.ucl.ac.uk}
	\and
	\alignauthor Deng Cai\\
		\affaddr{State Key Lab of CAD\&CG}\\
		\affaddr{Zhejiang University}\\
		\affaddr{100 Zijinggang Road}\\
		\affaddr{Hangzhou 310058, China}\\
		\email{dengcai@cad.zju.edu.cn}
	\alignauthor Jinsong Lu\\
		\affaddr{DEMS}\\
		\affaddr{Birkbeck, University of London}\\
		\affaddr{Malet Street}\\
		\affaddr{London WC1E 7HX, UK}\\
		\email{jingsong.lu@gmail.com}
}

\maketitle

\begin{abstract}
The ability of fast similarity search at large scale is of great importance to many Information Retrieval (IR) applications. A promising way to accelerate similarity search is semantic hashing which designs compact binary codes for a large number of documents so that semantically similar documents are mapped to similar codes (within a short Hamming distance). Although some recently proposed techniques are able to generate high-quality codes for documents known in advance, obtaining the codes for previously unseen documents remains to be a very challenging problem. In this paper, we emphasise this issue and propose a novel Self-Taught Hashing (STH) approach to semantic hashing: we first find the optimal $l$-bit binary codes for all documents in the given corpus via unsupervised learning, and then train $l$ classifiers via supervised learning to predict the $l$-bit code for any query document unseen before. Our experiments on three real-world text datasets show that the proposed approach using binarised Laplacian Eigenmap (LapEig) and linear Support Vector Machine (SVM) outperforms state-of-the-art techniques significantly.
\end{abstract}




\section{Introduction}
\label{sec:Introduction}

The problem of \emph{similarity search} (aka \emph{nearest neighbour search}) is: given a query document\footnote{In similarity search, a document is used as the query for retrieval, which is fundamentally different with the standard keyword search paradigm, e.g., in TREC.}, find its most similar documents from a very large document collection (corpus).
It is of great importance to many Information Retrieval (IR) \cite{Manning2008} applications, such as 
near-duplicate detection \cite{Henzinger2006}, 
plagiarism analysis \cite{Stein2007a}, 
collaborative filtering \cite{Koren2008},
caching \cite{Pandey2009}, and 
content-based multimedia retrieval \cite{Lew2006}.

Recently, with the rapid evolution of the Internet and the increased amounts of data to be processed, how to conduct fast similarity search at large scale has become an urgent research issue. 
A promising way to accelerate similarity search is \emph{semantic hashing} \cite{Salakhutdinov2009} which designs compact binary codes for a large number of documents so that semantically similar documents are mapped to similar codes (within a short Hamming distance). 
It is extremely fast to perform similarity search over such binary codes \cite{Stein2007}, because
\begin{itemize}
	\item the encoded data are highly compressed and thus can be loaded into the main memory; 
	\item the Hamming distance between two binary codes can be computed efficiently by using bit XOR operation and counting the number of set bits \cite{Knuth1997,Wegner1960}: an ordinary PC today would be able to do millions of Hamming distance computation in just a few milliseconds.
\end{itemize}
Furthermore, we usually just need to retrieve a small number of the most similar documents (i.e., nearest neighbours) for a given query document rather than computing its similarity to all documents in the collection. In such situations, we can simply return all the documents that are hashed into a tight Hamming ball centred around the binary code of the query document. 
For example, assuming that we use 4-bit binary codes, if the query document is represented as `\texttt{0000}', then we can just check this code as well as those 4 codes within one Hamming distance to it (i.e., having one bit difference with it) --- `\texttt{1000}', `\texttt{0100}', `\texttt{0010}', and `\texttt{0001}' --- and return the associated documents back.
It will also be easy to filter or re-rank the very small set of ``good'' documents (returned by semantic hashing) based on their full content, so as to further improve the retrieval effectiveness with just a little extra time \cite{Stein2007}.

In addition, similarity search serves as the basis of a classic non-parametric machine learning method, the k-Nearest-Neighbours (kNN) algorithm \cite{Mitchell1997}, for automated text categorisation \cite{Sebastiani2002} and so on. 
By enabling fast similarity search at large scale, semantic hashing makes it feasible to exploit ``the unreasonable effectiveness of data'' \cite{Halevy2009} to accomplish traditionally difficult tasks. 
For example, researchers recently achieved great success in scene completion and scene recognition using millions of images on the Web as training data \cite{Hays2007,Torralba2008}. 

\begin{figure*}[tb]
	\centering
		\includegraphics[scale=0.6]{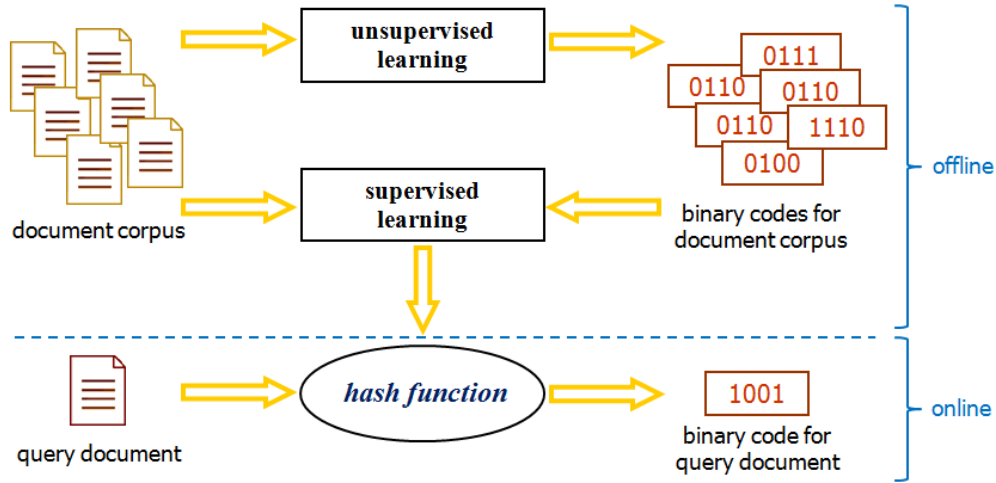}
	\caption{The proposed STH approach to semantic hashing.}
	\label{fig:STH}
\end{figure*}

Although some recently proposed techniques are able to generate high-quality codes for the documents known in advance, obtaining the codes for previously unseen documents remains to be a very challenging problem \cite{Stein2007}. 
Existing methods either have prohibitively high computational complexity or impose exceedingly restrictive assumptions about data distribution (see Section \ref{sec:SupervisedLearning}).
In this paper, we emphasise this issue and propose a novel Self-Taught Hashing (STH) approach to semantic hashing. As illustrated in Figure \ref{fig:STH}, we first find the optimal $l$-bit binary codes for all documents in the given corpus via unsupervised learning, and then train $l$ classifiers via supervised learning to predict the $l$-bit code for any query document unseen before. 

Our experiments on three real-world text datasets show that the proposed approach using binarised Laplacian Eigenmap (LapEig) \cite{Belkin2003} and linear Support Vector Machine (SVM) \cite{Joachims2002,Scholkopf2002} outperforms state-of-the-art techniques significantly, while maintaining a high running speed.

The rest of this paper is organised as follows. 
In Section \ref{sec:RelatedWork}, we review the related work.
In Section \ref{sec:Approach}, we present our approach in details.
In Section \ref{sec:Experiments}, we show the experimental results.
In Section \ref{sec:Conclusions}, we make conclusions.

\section{Related Work}
\label{sec:RelatedWork}

There has been extensive research on fast similarity search due to its central importance in many applications.
For a low-dimensional feature space, similarity search can be carried out efficiently with pre-built space-partitioning index structures (such as KD-tree) or data-partitioning index structures (such as R-tree) \cite{Cormen2001}. However, when the dimensionality of feature space is high (say $> 10$), similarity search aiming to return exact results cannot be done better than the naive method --- a linear scan of the entire collection \cite{Weber1998}. In the IR domain, documents are typically represented as feature vectors in a space of more than thousands of dimensions \cite{Manning2008}. Nevertheless, if the complete exactness of results is not really necessary, similarity search in a high-dimensional space can be dramatically speeded up by using hash-based methods which are purposefully designed to approximately answer queries in virtually constant time \cite{Stein2007}. 

Such hash-based methods for fast similarity search can be considered as a means for embedding high-dimensional feature vectors to a low-dimensional Hamming space (the set of all $2^l$ binary strings of length $l$), while retaining as much as possible the semantic similarity structure of data. Unlike standard dimensionality reduction techniques such as Latent Semantic Indexing (LSI) \cite{Berry1995,Deerwester1990} and Locality-Preserving Indexing (LPI) \cite{He2003,He2004}, hashing techniques map feature vectors to \emph{binary} codes, which is key to extremely fast similarity search (see Section \ref{sec:Introduction}). 
One possible way to get binary codes for text documents is to binarise the real-valued low-dimensional vectors (obtained from dimensionality reduction techniques like LSI) via thresholding \cite{Salakhutdinov2009}.
An improvement on binarised-LSI that directly optimises a Hamming distance based objective function, namely Laplacian Co-Hashing (LCH), has been proposed recently \cite{Zhang2010}.

The most well-known hashing technique that preserves similarity information is probably Locality-Sensitive Hashing (LSH) \cite{Andoni2006}. LSH simply employs random linear projections (followed by random thresholding) to map data points close in a  Euclidean space to similar codes. It is theoretically guaranteed that as the code length increases, the Hamming distance between two codes will asymptotically approach the Euclidean distance between their corresponding data points. However, since the design of hash functions for LSH is \emph{data-oblivious}, LSH may lead to quite inefficient (long) codes in practice \cite{Salakhutdinov2009,Weiss2008}.

Several recently proposed hashing techniques attempt to overcome this problem by finding good \emph{data-aware} hash functions through machine learning.
In \cite{Salakhutdinov2009}, the authors proposed to use stacked Restricted Boltzmann Machine (RBM) \cite{Hinton2006,Hinton2006a}, and showed that it was indeed able to generate compact binary codes to accelerate document retrieval. 
Researchers have also tried the boosting approach to Similarity Sensitive Coding (SSC) \cite{Shakhnarovich2003} and Forgiving Hashing (FgH) \cite{Baluja2008} --- they first train AdaBoost \cite{Schapire2003} classifiers with similar pairs of items as positive examples (and also non-similar pairs of items as negative examples in SCC), and then take the output of all (decision stump) weak learners on a given document as its binary code.
In \cite{Torralba2008}, both stacked-RBM and boosting-SSC were found to work significantly better and faster than LSH when applied to a database containing tens of millions of images.
In \cite{Weiss2008}, a new technique called Spectral Hashing (SpH) was proposed. It has demonstrated significant improvements over LSH, stacked-RBM and boosting-SSC in terms of the number of bits required to find good similar items.
There is some resemblance between the first step of SpH and the unsupervised learning stage of our STH approach, because both are related to spectral graph partitioning \cite{Chung1997,Hagen1992,Shi2000}. Nevertheless, we use a different spectral method and take a different way to address the \emph{entropy maximising} criterion (see Section \ref{sec:UnsupervisedLearning}). More importantly, in order to process query documents, SpH has to assume that the data are uniformly distributed in a hyper-rectangle, which is apparently very restrictive. In contrast, our proposed STH approach can work with any data distribution and it is much more flexible (see Section \ref{sec:SupervisedLearning}). The superiority of STH to SpH has been confirmed by our experimental results (see Section \ref{sec:Experiments}). 

A somewhat related, but different, line of research is to use hashing representations for machine learning \cite{Shi2009,Weinberger2009}. The objective of such techniques is to accelerate complex learning algorithms, but not similarity search. Our work is basically the other way around.

\section{Approach}
\label{sec:Approach}

The proposed Self-Taught Hashing (STH) approach to semantic hashing is a \emph{general} learning framework that consists of two distinct stages, as illustrated in Figure \ref{fig:STH}. 
We call the approach ``self-taught'' because the hash function is learnt from the data that are auto-labelled by itself in the previous stage\footnote{It is, however, worth noticing that the term ``self-taught learning'' has been mentioned in \cite{Raina2007} where the intention was to describe a strategy for transfer learning based on sparse coding, whereas in this
paper the term has a rather different meaning.}. 

\subsection{Stage 1:\\Unsupervised Learning of Binary Codes}
\label{sec:UnsupervisedLearning}

Given a collection of $n$ documents which are represented as $m$-dimensional vectors $\{\mathbf{x}_i\}_{i=1}^n \subset \mathbb{R}^m$.
Let $X$ denote the $m \times n$ term-document matrix: $[\mathbf{x}_1,\ldots,\mathbf{x}_n]$.
Suppose that the desired length of code is $l$ bits. 
We use $\mathbf{y}_i \in \{-1,+1\}^l$ to represent the binary code for document vector $\mathbf{x}_i$, where the $p$-th element of $\mathbf{y}_i$, i.e., $y_i^{(p)}$, is $+1$ if the $p$-th bit of code is on, or $-1$ otherwise.
Let $Y$ denote the $n \times l$ matrix whose $i$-th row is the code for the $i$-th document, i.e., $[\mathbf{y}_1,\ldots,\mathbf{y}_n]^T$. 

A ``good'' semantic hashing should be \textbf{\emph{similarity preserving}} to ensure effectiveness. That is to say, semantically similar documents should be mapped to similar codes within a short Hamming distance.

Unlike the existing approaches (such as SpH \cite{Weiss2008}) that aim to preserve the \emph{global} similarity structure of all document pairs, we focus on the \emph{local} similarity structure, i.e., $k$-nearest-neighbourhood, for each document. Since IR applications usually put emphasis on a small number of most similar documents for a given query document \cite{Manning2008}, preserving the global similarity structure is not only unnecessary but also likely to be sub-optimal for our problem. 
Therefore, using the \emph{cosine similarity}\footnote{Our approach can work with any legitimate similarity measure, though we focus on cosine similarity in this paper.} \cite{Manning2008}, we construct our $n \times n$ \emph{local} similarity matrix $W$ as 
\begin{equation}
\label{eq:SimilarityMatrix}
W_{ij} = \left\{ 
\begin{array}{cl}
\frac{\mathbf{x}_i^T\mathbf{x}_j}{\|\mathbf{x}_i\|\cdot\|\mathbf{x}_j\|} &
 \qquad \text{if } \mathbf{x}_i \in N_k(\mathbf{x}_j) \text{ or }
 \mathbf{x}_j \in N_k(\mathbf{x}_i) \\
0 & \qquad\text{otherwise}
\end{array}
\right.
\end{equation}
where $N_k(\mathbf{x})$ represents the set of $k$-nearest-neighbours of document $\mathbf{x}$. In other words, $W$ is the adjacency matrix of the $k$-nearest-neighbours graph for the given corpus \cite{Belkin2003}. A by-product of focusing on such a local similarity structure instead of the global one is that $W$ becomes a sparse matrix. This not only leads to much lower storage overhead, but also brings a significant reduction to the computational complexity of subsequent operations.
Furthermore, we introduce a diagonal $n \times n$ matrix $D$ whose entries are given by $D_{ii} = \sum_{j=1}^n W_{ij}$. The matrix $D$ provides a natural measure of document importance: the bigger the value of $D_{ii}$ is, the more ``important'' is the document $\mathbf{x}_i$ as its neighbours are strongly connected to it \cite{Belkin2003}. 

The Hamming distance between two binary codes $\mathbf{y}_i$ and $\mathbf{y}_j$ (corresponding to documents $\mathbf{x}_i$ and $\mathbf{x}_j$) is given by the number of bits that are different between them, which can be calculated as $\frac{1}{4} \|\mathbf{y}_i-\mathbf{y}_j\|^2$. 
To meet the \emph{similarity preserving} criterion, we seek to minimise the weighted average Hamming distance (as in SpH \cite{Weiss2008})
\begin{equation}
\label{eq:ObjectiveFunction}
\frac{1}{4} \sum_{i=1}^n \sum_{j=1}^n W_{ij} \|\mathbf{y}_i-\mathbf{y}_j\|^2
\end{equation}
because it incurs a heavy penalty if two similar documents are mapped far apart.
After some simple mathematical transformation, the above objective function can be rewritten in matrix form as $\frac{1}{4} \Tr(Y^T L Y)$, where $L = D-W$ is the \emph{graph Laplacian} \cite{Chung1997}, and $\Tr(\cdot)$ means the matrix trace. 

We found the above objective function \eqref{eq:ObjectiveFunction} actually proportional to that of a well-known manifold learning algorithm, Laplacian Eigenmap (LapEig) \cite{Belkin2003}, except that LapEig does not have the constraint $\mathbf{y}_i \in \{-1,+1\}^l$. So, if we relax this discreteness condition but just keep the \emph{similarity preserving} requirement, we can get the optimal $l$-dimensional real-valued vector $\tilde{\mathbf{y}}_i$ to represent each document $\mathbf{x}_i$ by solving the following LapEig problem:
\begin{align}
\label{eq:LaplacianEigenmap}
\argmin_{\tilde{Y}} \qquad & \Tr(\tilde{Y}^T L \tilde{Y}) \\
\text{subject to}
\qquad & \tilde{Y}^T D \tilde{Y} = I \nonumber \\
\qquad & \tilde{Y}^T D \mathbf{1} = \mathbf{0} \nonumber
\end{align}
where $\Tr(\tilde{Y}^T L \tilde{Y})$ gives the real relaxation of the weighted average Hamming distance $\Tr(Y^T L Y)$, and the two constraints prevent the collapse into a subspace of dimension less than $l$.
The solution of this optimisation problem is given by $\tilde{Y} = [\mathbf{v}_1,\ldots,\mathbf{v}_l]$ whose columns are the $l$ eigenvectors corresponding to the smallest eigenvalues of the following generalised eigenvalue problem (except the trivial eigenvalue 0):
\begin{equation}
\label{eq:GeneralisedEigenvalueProblem}
L \mathbf{v} = \lambda D \mathbf{v}
\end{equation}

The above LapEig formulation \eqref{eq:LaplacianEigenmap} may look similar to the first step of SpH \cite{Weiss2008}. This is because SpH is motivated by a spectral graph partitioning method \emph{ratio-cut} \cite{Hagen1992}, while LapEig is closely connected to another spectral graph partitioning method \emph{normalised-cut} \cite{Shi2000}. Many independent studies have shown that normalised-cut has better theoretical properties and empirical performances than ratio-cut \cite{Chung1997,Shi2000}. 

We now convert the above $l$-dimensional real-valued vectors $\tilde{\mathbf{y}}_1,\ldots,\tilde{\mathbf{y}}_n$ into binary codes via thresholding: if the $p$-th element of $\tilde{\mathbf{y}}_i$ is larger than the specified threshold, $y_i^{(p)} = +1$ (i.e., the $p$-th bit of the $i$-th code is on); otherwise, $y_i^{(p)} = -1$ (i.e., the $p$-th bit of the $i$-th code is off).

A ``good'' semantic hashing should also be \textbf{\emph{entropy maximising}} to ensure efficiency, as pointed out by \cite{Baluja2008}. According to the \emph{information theory} \cite{Shannon1948}: the maximal entropy of a source alphabet is attained by having a uniform probability distribution. If the entropy of codes over the corpus is small, it means that documents are mapped to only a small number of codes (hash bins), thereby rendering the hash table inefficient.
To meet this \emph{entropy maximising} criterion, we set the threshold for binarising $\tilde{y}_1^{(p)},\ldots,\tilde{y}_n^{(p)}$ to be the \emph{median} value of $\mathbf{v}_p$. In this way, the $p$-th bit will be on for half of the corpus and off for the other half. Furthermore, as the eigenvectors $\mathbf{v}_1,\ldots,\mathbf{v}_l$ given by LapEig are orthogonal to each other, different bits $y^{(1)},\ldots,y^{(l)}$ in the generated binary codes will be uncorrelated. Therefore this thresholding method gives each distinct binary code roughly equal probability of occurring in the document collection, thus achieves the best utilisation of the hash table.

\subsection{Stage 2:\\Supervised Learning of Hash Function}
\label{sec:SupervisedLearning}

Mapping all documents in the given corpus to binary codes does not completely solve the problem of semantic hashing, because we also need to know how to obtain the binary codes for query documents, i.e., new documents that are unseen before. 
This problem, called \emph{out-of-sample extension} in manifold learning, is often addressed using the Nystrom method \cite{Belongie2002,Drineas2005}. However, calculating the Nystrom extension of a new document is as computationally expensive as an exhaustive similarity search over the corpus (that may contain millions of documents), which makes it impractical for semantic hashing.
In LPI \cite{He2003,He2004}, LapEig \cite{Belkin2003} is extended to deal with new samples by approximating a linear function to the embedding of LapEig. However, the computational complexity of LPI is very high because its learning algorithm involves eigen-decompositions of two large dense matrices. It is infeasible to apply LPI if the given training corpus is large.
In SpH \cite{Weiss2008}, new samples are handled by utilising the latest results on the convergence of graph Laplacian eigenvectors to the Laplace-Beltrami eigenfunctions of manifolds. It can achieve both fast learning and fast prediction, but it relies on a very restrictive assumption that the data are uniformly distributed in a hyper-rectangle.

Overcoming the limitations of the above techniques \cite{Belongie2002,Drineas2005,He2003,He2004,Weiss2008}, this paper proposes a novel method to compute the binary codes for query documents by considering it as a supervised\footnote{Since in the second stage, the supervised learning algorithm uses only the \emph{pseudo}-labels input from the previous unsupervised learning stage, the entire STH approach remains to be unsupervised.} learning problem: we think of each bit $y_i^{(p)} \in \{+1,-1\}$ in the binary code for document $\mathbf{x}_i$ as a binary class label (class-``on'' or class-``off'') for that document, and train a binary classifier $y^{(p)} = f^{(p)}(\mathbf{x})$ on the given corpus that has already been ``labelled'' by the above binarised-LapEig method, then we can use the learned binary classifiers $f^{(1)},\ldots,f^{(l)}$ to predict the $l$-bit binary code $y^{(1)},\ldots,y^{(l)}$ for any query document $\mathbf{x}$. 
As mentioned in the previous section, different bits $y^{(1)},\ldots,y^{(l)}$ in the generated binary codes are uncorrelated. Hence there is no redundancy among the binary classifiers $f^{(1)},\ldots,f^{(l)}$, and they can also be trained independently.

In this paper, we choose to use the Support Vector Machine (SVM) \cite{Joachims2002,Scholkopf2002} algorithm to train these binary classifiers. SVM in its simplest form, linear SVM $f(\mathbf{x}) = \sgn(\mathbf{w}^T \mathbf{x})$ consistently provides state-of-the-art performance for text classification tasks \cite{Dumais1998,Joachims1998,Yang1999}. 
Given the documents $\mathbf{x}_1,\ldots,\mathbf{x}_n$ together with their {\emph{self-taught} binary labels for the $p$-th bit $y_1^{(p)},\ldots,y_n^{(p)}$, the corresponding linear SVM can be trained by solving the following quadratic optimisation problem
\begin{align}
\label{eq:LinearSVM}
\argmin_{\mathbf{w}, \xi_i \geq 0} \qquad & \frac{1}{2} {\mathbf{w}^T \mathbf{w}} + \frac{C}{n} {\sum_{i=1}^{n} \xi_i} \\
\text{subject to}
\qquad & \forall_{i=1}^{n}: y_i^{(p)} \mathbf{w}^T \mathbf{x}_i \geq 1 - \xi_i \nonumber
\end{align}
A notable advantage of using SVM classifiers here is that we can easily achieve non-linear mappings if necessary by plugging in non-linear kernels \cite{Scholkopf2002}, though we do not explore this potential in this paper.

\subsection{Summary of Approach}
\label{sec:Summary}

We name the above proposed two-stage approach Self-Taught Hashing (STH). In this paper, we choose binarised-LapEig \cite{Belkin2003} for the unsupervised learning stage and linear-SVM \cite{Joachims2002,Scholkopf2002} for the supervised learning stage, but obviously it is possible to use other machine learning algorithms.

The \emph{learning} process of STH for a given corpus can be summarized as follows.
\begin{enumerate}
	\item \textbf{unsupervised learning of binary codes}:
	\begin{itemize}
		\item construct the $k$-nearest-neighbours graph for the given corpus;
		\item embed the documents in an $l$-dimensional space through LapEig \eqref{eq:GeneralisedEigenvalueProblem} to get an $l$-dimensional real-valued vector for each document;
		\item obtain an $l$-bit binary code for each document via thresholding the above vectors at their median point, and then take each bit as a binary class label for that document;
	\end{itemize}
	\item \textbf{supervised learning of hash function}:
	\begin{itemize}
		\item train $l$ SVM classifiers \eqref{eq:LinearSVM} based on the given corpus that has been ``labelled'' as above.
	\end{itemize}
\end{enumerate}
Let $s$ denote the average number of non-zero features per document. 
In the first stage, constructing the $k$-nearest-neighbours graph takes $\BigO(n^2 s + n^2 k)$ time using the selection algorithm \cite{Cormen2001}, solving the LapEig problem \eqref{eq:GeneralisedEigenvalueProblem} takes $\BigO(l n k t)$ time using the Lanczos algorithm \cite{Golub1996} of $t$ iterations (the value of $t$ is usually quite small), and the median-based binarisation takes $\BigO(l n)$ time again using the selection algorithm \cite{Cormen2001}.
In the second stage, thanks to the recent advances in large-scale optimisation, each of the $l$ linear SVM classifiers can be trained in $\BigO(s n)$ time or even less \cite{Joachims2006,Hsieh2008}, so all training can be done in $\BigO(l s n)$ time. 
Both the value of $l$ and the value of $k$ can be regarded as small constants, as usually a short code length is desirable and just a few nearest neighbours are needed. For example, $l \leq 64$ and $k = 25$ in our experiments (see Section \ref{sec:Experiments}).
Therefore the overall computational complexity of the learning process is roughly quadratic to the number of documents in the corpus while linear to the average size of the documents in the corpus.

The \emph{predicting} process of STH for a given query document is simply to classify the query document using those $l$ learned classifiers and then assemble the output $l$ binary labels into an $l$-bit binary code.
For linear SVM, classifying a document only requires one dot-product operation between two vectors, the aggregated support vector and the document vector (with $s'$ non-zero features), which can be done quickly in $\BigO(s')$ time.
Therefore the overall computational complexity of the prediction process for each query document is linear to the size of the query document.

\section{Experiments}
\label{sec:Experiments}

We now empirically evaluate our proposed STH approach (using binarised-LapEig and linear-SVM), and compare its performance with binarised-LSI \cite{Salakhutdinov2009}, LCH \cite{Zhang2010}, and SpH \cite{Weiss2008} that represents the state of the art (see Section \ref{sec:RelatedWork}).

In the following STH experiments, the parameter $k=25$ when constructing the $k$-nearest-neighbours graph for LapEig\footnote{In principle, the value of $k$ for LapEig should be set to the desired number of original nearest neighbours to be retrieved (see Section \ref{sec:Evaluation}).}, and the SVM implementation is from LIBLINEAR \cite{Fan2008} with the default parameter values\footnote{It is not necessary to fine tune the SVM parameters (such as $C$) because it has already worked very well with its default parameter values.}.

\subsection{Data}
\label{sec:Data}

We have conducted experiments on three publicly available real-world text datasets:
Reuters21578\footnote{http://www.daviddlewis.com/resources/testcollections/ reuters21578/}, 
20Newsgroups\footnote{http://people.csail.mit.edu/jrennie/20Newsgroups/} and 
TDT2\footnote{http://www.nist.gov/speech/tests/tdt/tdt98/index.htm}. 

The Reuters21578 corpus is a collection of documents that appeared on Reuters newswire in 1987. 
It contains 21578 documents in 135 categories. 
In our experiments, those documents appearing in more than one category were discarded, and only the largest 10 categories were kept, thus leaving us with 7285 documents in total.
We use the ModeApte split here which gives 5228 (72\%) documents for training and 2057 (28\%) documents for testing.

The 20Newsgroups corpus was collected and originally used for document categorisation by Lang \cite{Lang1995}. We use the popular `bydate' version which contains 18846 documents, evenly distributed across 20 categories.
The time-based split leads to 11314 (60\%) documents for training and 7532 (40\%) documents for testing. 

The TDT2 (NIST Topic Detection and Tracking) corpus consists of data collected during the first half of 1998 and taken from 6 sources, including 2 newswires (APW, NYT), 2 radio programs (VOA, PRI) and 2 television programs (CNN, ABC). 
It consists of 11201 on-topic documents which are classified into 96 semantic categories. 
In our experiments, those documents appearing in more than one category were discarded, and only the largest 30 categories were kept, thus leaving us with 9394 documents in total.
We randomly selected 5597 (60\%) documents for training and 3797 (40\%) documents for testing. The averaged performance based on 10 such random selections is reported in this paper. 

All the above datasets have been pre-processed by stop-word removal, Porter stemming, and TF-IDF weighting \cite{Manning2008}. 

For the purpose of reproducibility, we shall make the datasets and code used in our experiments publicly available at the first author's homepage upon paper publication.

\subsection{Evaluation}
\label{sec:Evaluation}

Given a dataset, we use each document in the test set as a query to retrieve documents in the training set within a specified Hamming distance, and then compute standard retrieval performance measures: \emph{precision}, \emph{recall}, and their harmonic mean ($F_1$ measure) \cite{Manning2008}.
\begin{align}
precision &= \frac{\text{the number of retrieved relevant documents}}{\text{the number of all retrieved documents}} \\
recall &= \frac{\text{the number of retrieved relevant documents}}{\text{the number of all relevant documents}}
\end{align}
The reported performance scores in the following Section are averaged over all test queries in the dataset.

To determine whether a retrieved document is ``relevant'' to the given query document, we adopt the following two evaluation methodologies:
\begin{enumerate}
	\item \textbf{retrieving original nearest neighbours} --- the $k$ most similar documents, i.e., nearest neighbours, in the original vector space are considered as the ground-truth relevant documents ($k=25$ in our experiments);
	\item \textbf{retrieving same-topic documents} --- the documents on the same topic, i.e., in the same category, are considered as the ground-truth relevant documents.
\end{enumerate}
The former methodology is used in \cite{Weiss2008}\footnote{Actually only precision is used in \cite{Weiss2008}, which is appropriate for their application of pattern recognition but obviously insufficient from the IR perspective. Due to this difference in performance measurement, their results are not directly comparable with ours.}, 
while the latter methodology is used in \cite{Salakhutdinov2009}. 
In our opinion, these two methodologies emphasise different aspects of semantic hashing, and thus are suitable for different target IR applications. Therefore we use both of them in this paper. 

The absolute performance scores of STH are not as important as how they compare with those of other semantic hashing techniques. 
As previously mentioned in Section \ref{sec:Introduction}, if necessary, we can always spend a little extra time to filter or re-rank the similarity search results based on their full content, thus achieve higher performance scores \cite{Stein2007}.

\subsection{Results}
\label{sec:Results}

Figure \ref{fig:trueF1} and Figure \ref{fig:cateF1} show the $F_1$ measure of STH for retrieving original nearest neighbours and same-topic documents respectively\footnote{The $F_1$ measure scores reported here should not be directly compared with those in text categorisation papers, as we are addressing a very different problem even though the same datasets may have been used for experimentation.}. 
We vary the code length from 4-bit to 64-bit and also the Hamming ball radius (i.e., the maximum Hamming distance between any retrieved document and the query document) from 0 to 3, in order to show their influences on the retrieval performance.
It can be seen that when the code length increases, STH is able to achieve a higher $F_1$ measure (using a bigger Hamming ball radius). However, longer binary codes demand more memory and a bigger Hamming ball radius requires more computation. The optimal trade-off between effectiveness and efficiency can be found by using a validation set of query documents.    

\begin{figure*}[p]
	\centering
	\subfigure[Reuters21578]{
		\includegraphics[width=0.3\textwidth]{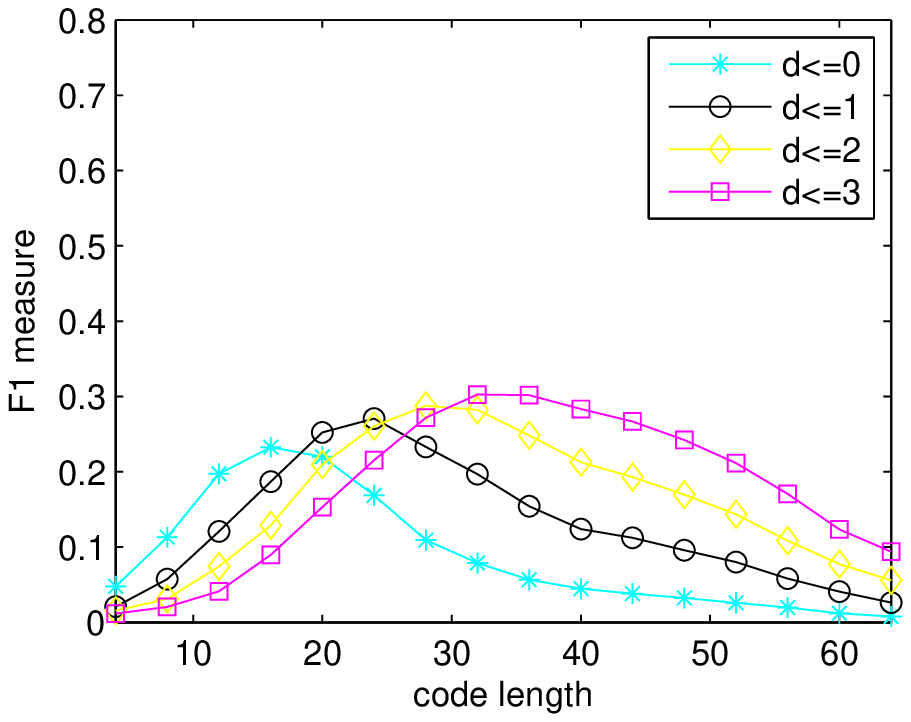}
		\label{fig:trueF1_Reuters21578_STH}
	}
	\subfigure[20Newsgroups]{
		\includegraphics[width=0.3\textwidth]{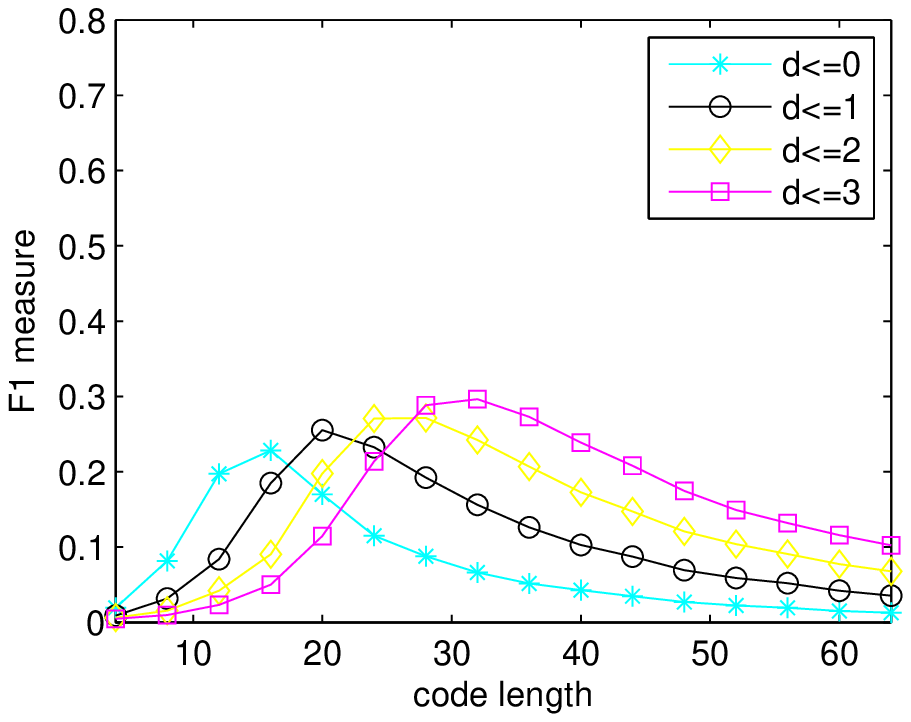}
		\label{fig:trueF1_20Newsgroups_STH}
	}
	\subfigure[TDT2]{
		\includegraphics[width=0.3\textwidth]{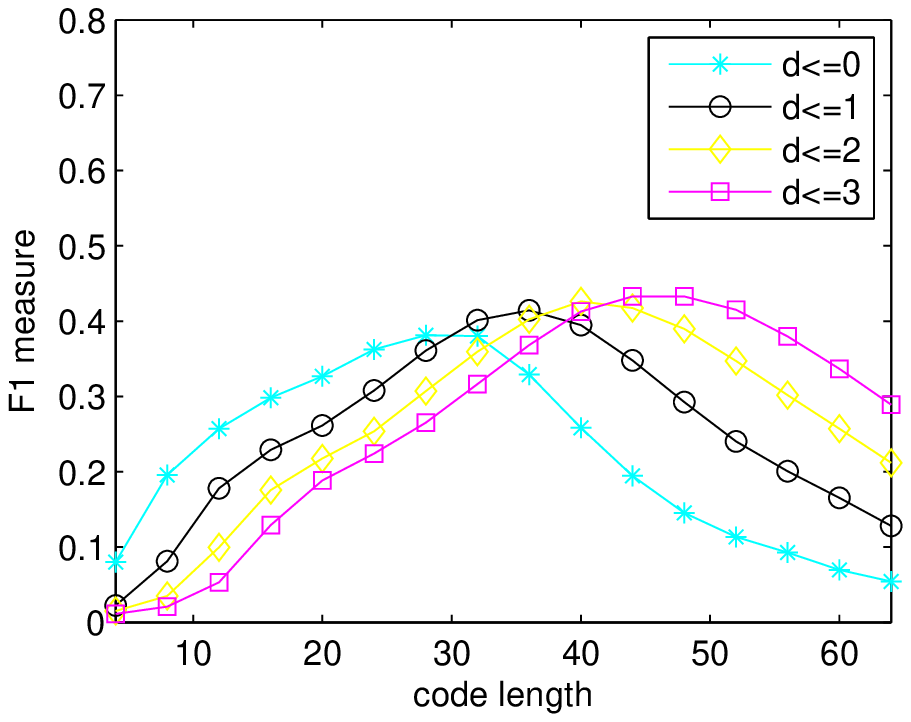}
		\label{fig:trueF1_TDT2_STH}
	}
	\caption{The $F_1$ measure of STH for retrieving original nearest neighbours.}
	\label{fig:trueF1}
\end{figure*}

\begin{figure*}[p]
	\centering
	\subfigure[Reuters21578]{
		\includegraphics[width=0.3\textwidth]{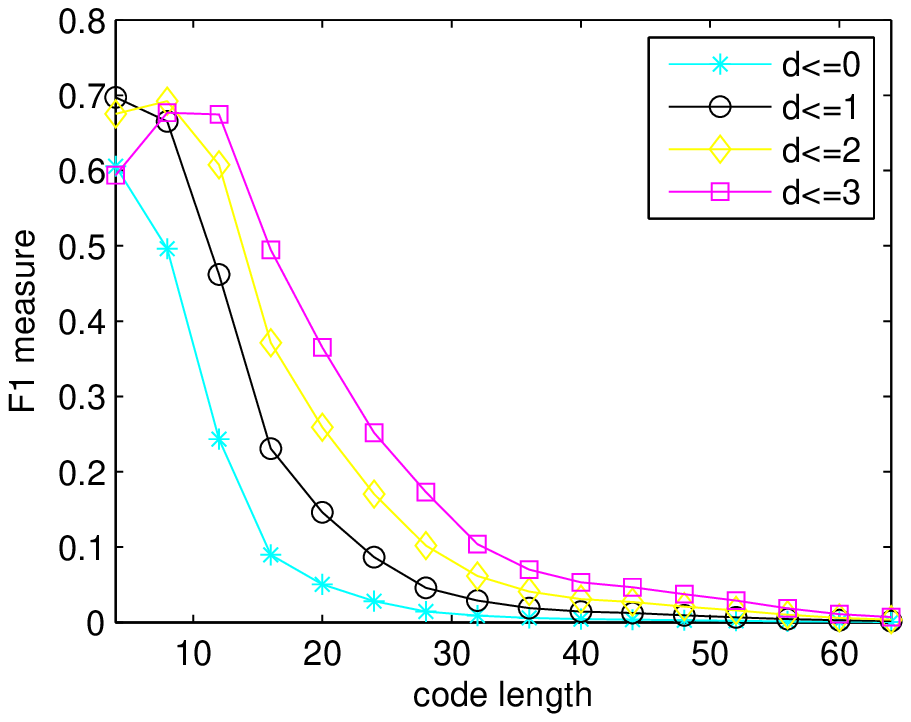}
		\label{fig:cateF1_Reuters21578_STH}
	}
	\subfigure[20Newsgroups]{
		\includegraphics[width=0.3\textwidth]{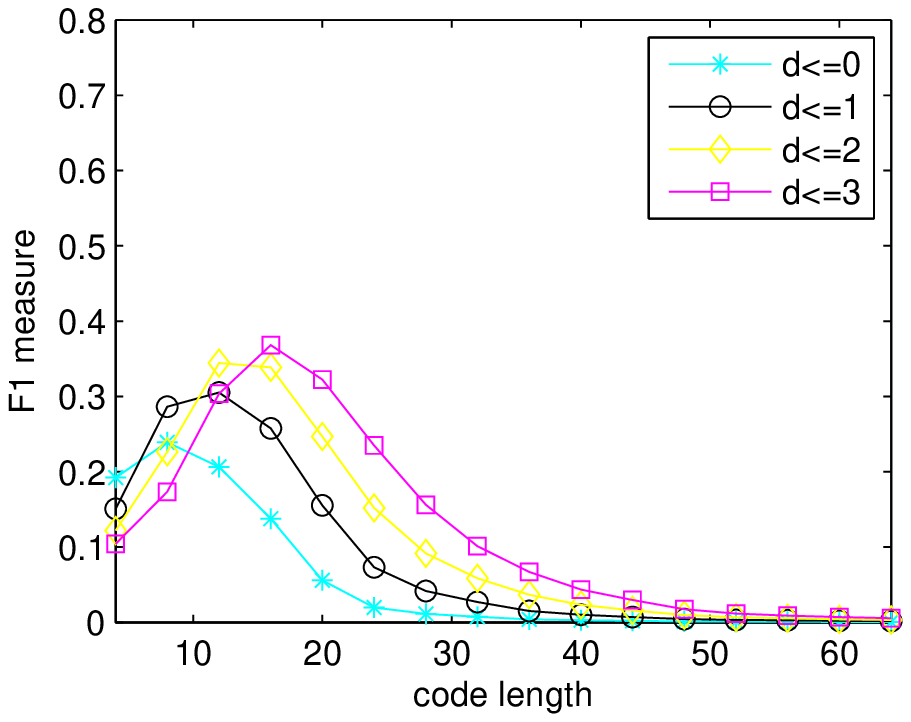}
		\label{fig:cateF1_20Newsgroups_STH}
	}
	\subfigure[TDT2]{
		\includegraphics[width=0.3\textwidth]{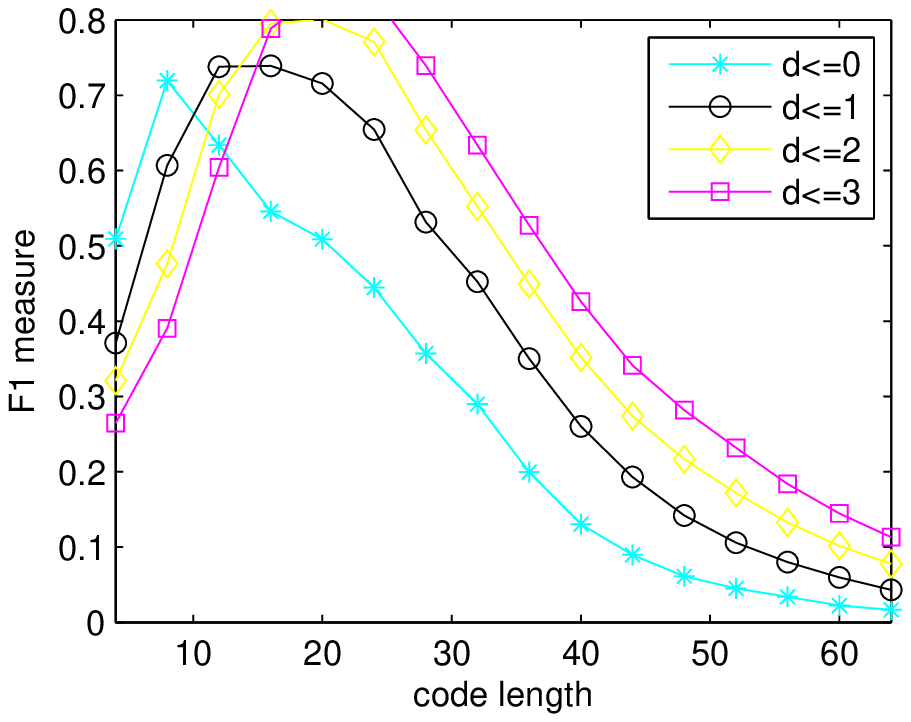}
		\label{fig:cateF1_TDT2_STH}
	}
	\caption{The $F_1$ measure of STH for retrieving same-topic documents.}
	\label{fig:cateF1}
\end{figure*}

Figure \ref{fig:truePR} and Figure \ref{fig:catePR} compare STH with several other typical semantic hashing methods in terms of their \emph{precision-recall curves} (created by varying the code length from 4-bit to 64-bit while fixing the Hamming ball radius at $1$), for retrieving original nearest neighbours and same-topic documents respectively\footnote{Although we could achieve higher retrieval performance by utilising a bigger Hamming ball radius (e.g., 4), a large number of binary codes (e.g., $C_{64}^4 = 635376$ for 64-bit codes) would need to be checked for each query and then the efficiency gain brought by semantic hashing would diminish.}.
It is clear that on all datasets and under both evaluation methodologies, STH outperforms binarised-LSI, LCH, and the state-of-the-art technique SpH (that has already been shown to work much better than LSH \cite{Andoni2006}, stacked-RBM\footnote{For example, on the 20Newsgroups dataset, stacked-RBM achieves a maximum of $F_1 = 0.276$ for retrieving same-topic documents with 128-bit codes, while the same level of performance can be obtained using our STH approach with just 8-bit codes.} \cite{Salakhutdinov2009} and boosting-SSC \cite{Shakhnarovich2003}).
Using 16-bit codes and Hamming ball radius $1$, the performance improvements are all statistically significant ($P$ value $<0.01$) according to one-sided micro sign test ($s$-test) \cite{Yang1999}.

\begin{figure*}[p]
	\centering
	\subfigure[Reuters21578]{
		\includegraphics[width=0.3\textwidth]{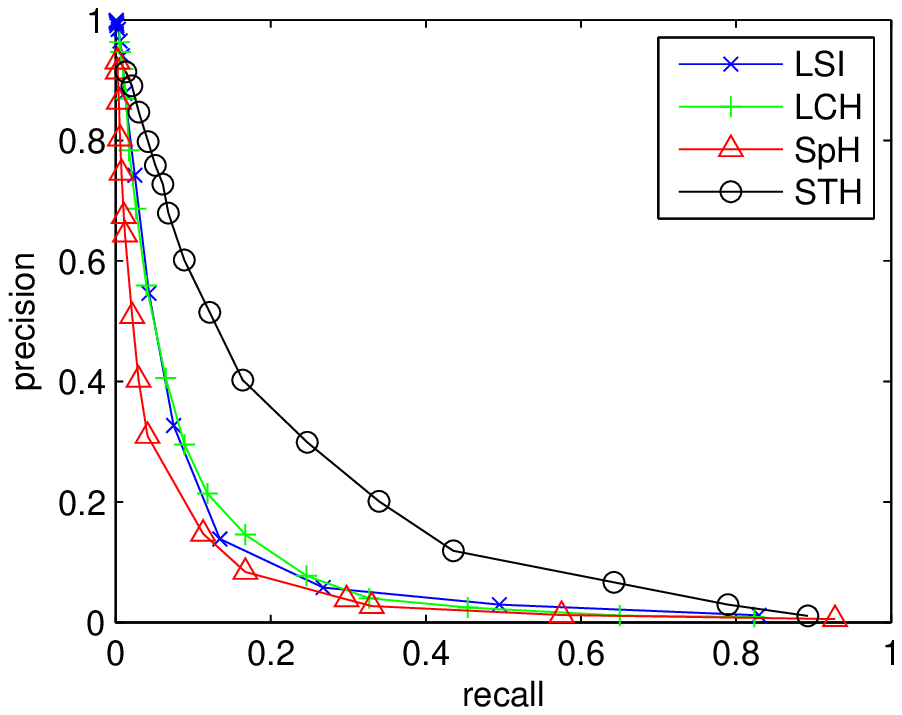}
		\label{fig:truePR_Reuters21578}
	}
	\subfigure[20Newsgroups]{
		\includegraphics[width=0.3\textwidth]{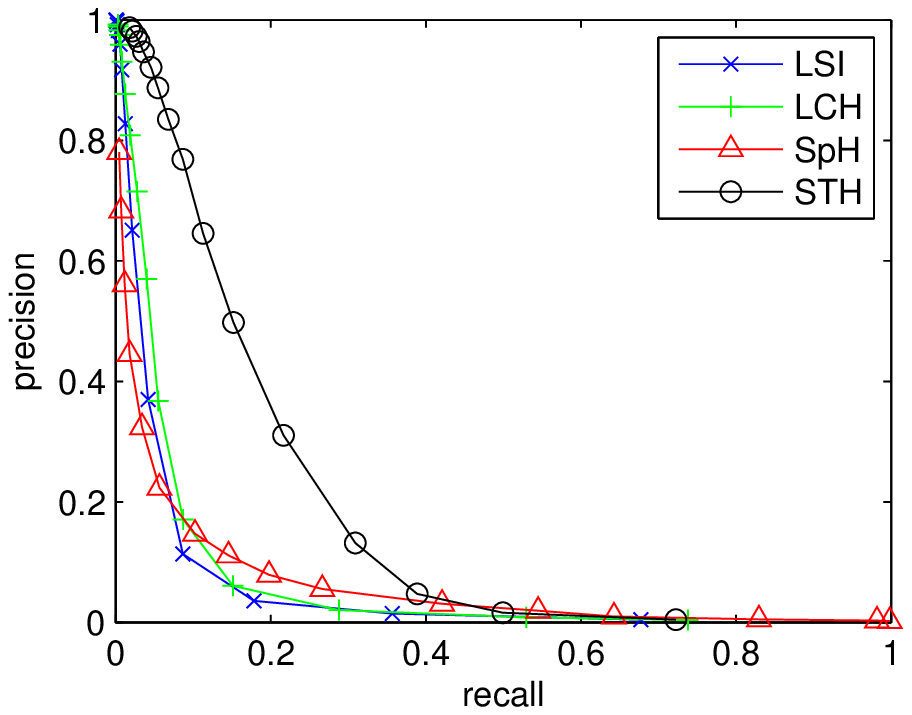}
		\label{fig:truePR_20Newsgroups}
	}
	\subfigure[TDT2]{
		\includegraphics[width=0.3\textwidth]{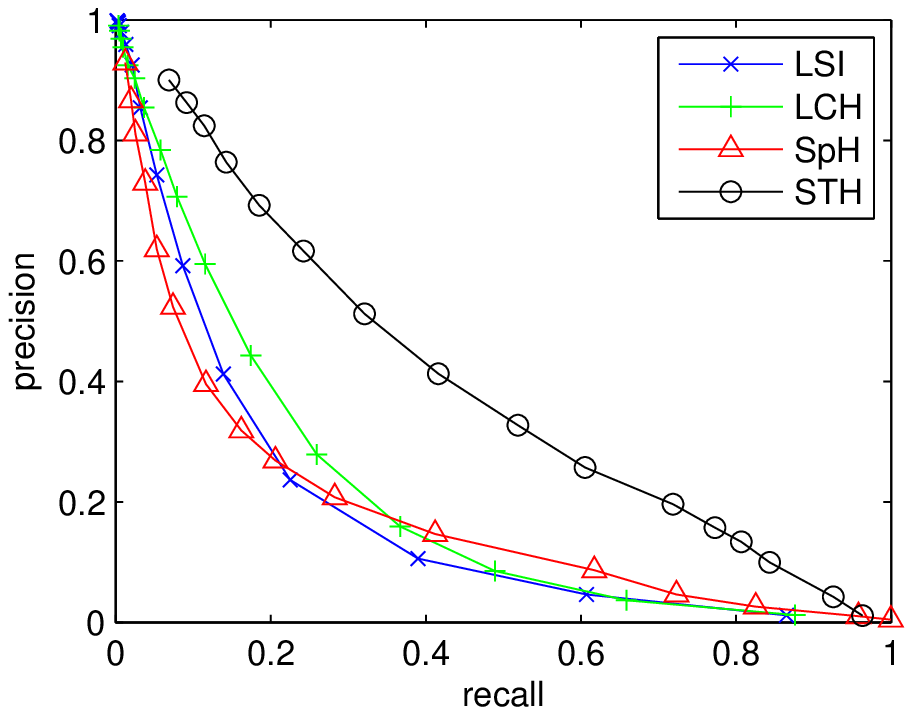}
		\label{fig:truePR_TDT2}
	}
	\caption{The precision-recall curve for retrieving original nearest neighbours.}
	\label{fig:truePR}
\end{figure*}

\begin{figure*}[p]
	\centering
	\subfigure[Reuters21578]{
		\includegraphics[width=0.3\textwidth]{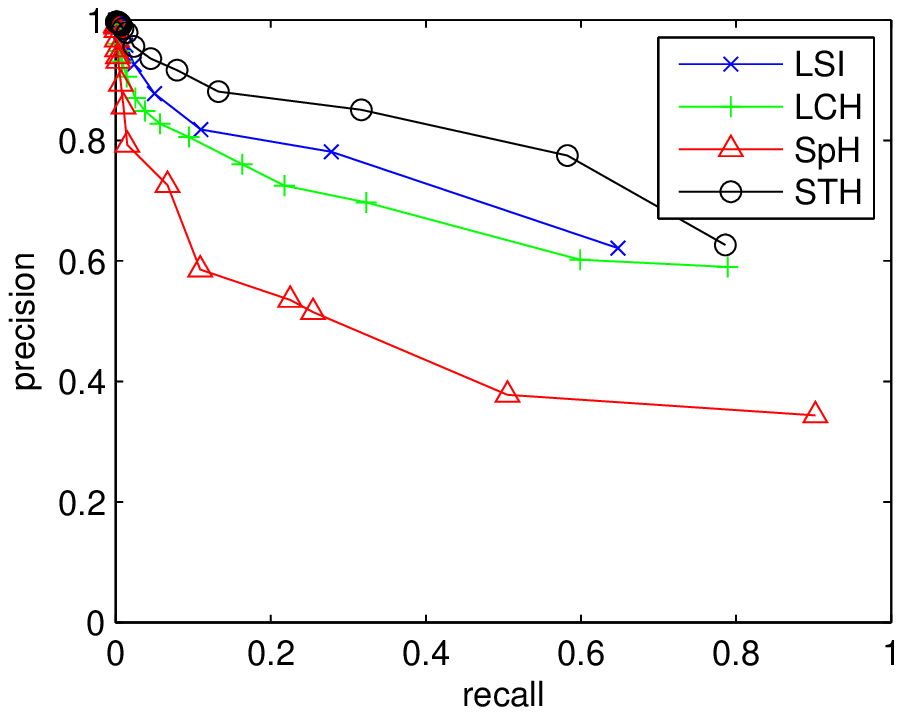}
		\label{fig:catePR_Reuters21578}
	}
	\subfigure[20Newsgroups]{
		\includegraphics[width=0.3\textwidth]{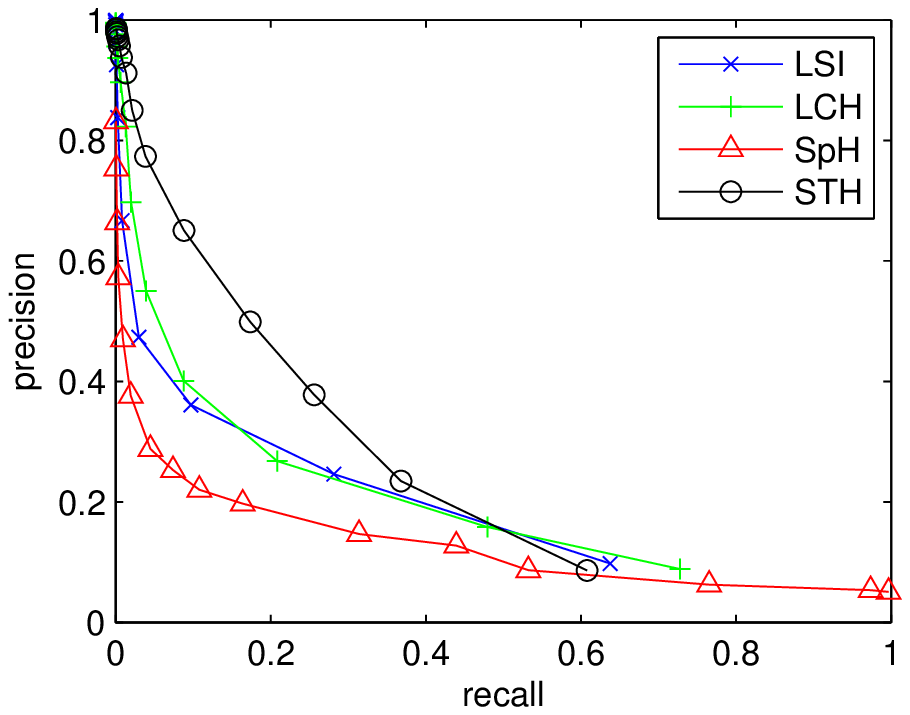}
		\label{fig:catePR_20Newsgroups}
	}
	\subfigure[TDT2]{
		\includegraphics[width=0.3\textwidth]{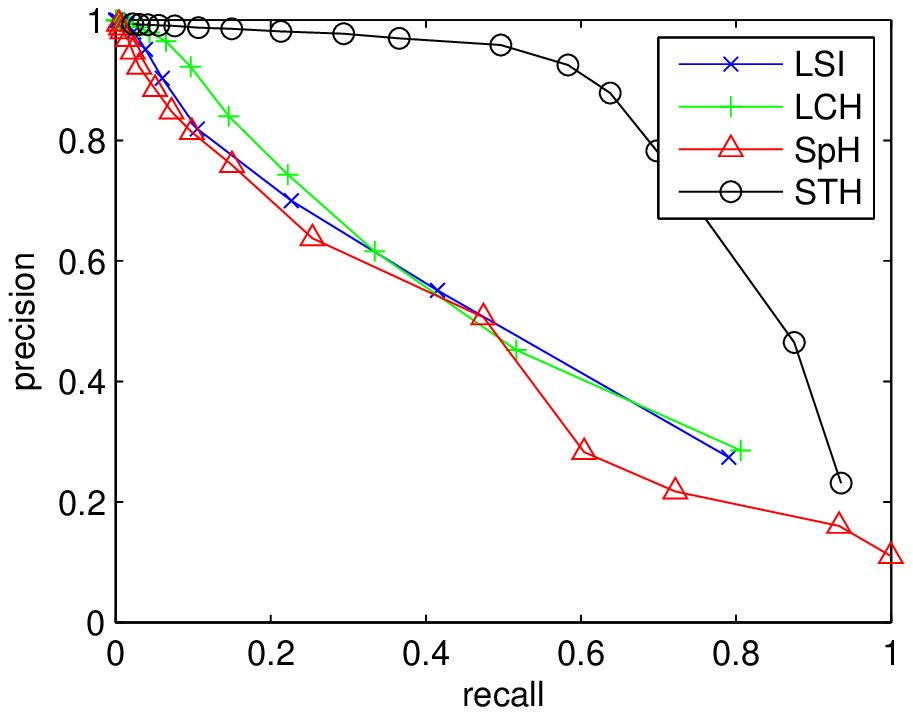}
		\label{fig:catePR_TDT2}
	}
	\caption{The precision-recall curve for retrieving same-topic documents.}
	\label{fig:catePR}
\end{figure*}

We think the superior performance of STH is due to two reasons: 
\begin{itemize}
	\item the binary codes produced by binarised-LapEig effectively preserve the semantic similarity structure while maximising the entropy of the hash table;
	\item the maximum-margin hyperplane produced by linear-SVM ensures high generalisation ability \cite{Scholkopf2002}. 
\end{itemize}

We have also examined the approximation errors accumulated in each step of STH (see Section \ref{sec:Summary}). Our anatomy reveals that almost all approximation errors come from the dimensionality reduction step using LapEig. 
However, LapEig does work better than alternative methods (such as LSI) for this step in our experiments, and it is a well-known hard problem to accurately detect the (intrinsic) dimensionality of data or effectively reduce the dimensionality of data. 
The median-based binarisation and SVM-based out-of-sample extension both work perfectly incurring little approximation errors.

The proposed STH approach (using binarised-LapEig and linear-SVM) to semantic hashing is pretty fast: on an ordinary PC with Intel Pentium4 3.00GHz CPU and 2GB RAM, our Matlab implementation of 64-bit STH takes approximately 0.0165 second per document for training (which is about 10 times faster than SpH), and 0.0007 second per document for prediction.

\section{Conclusions}
\label{sec:Conclusions}

The main contribution of this paper is a novel Self-Taught Hashing (STH) approach to semantic hashing for fast similarity search. By decomposing the problem of finding small codes for large data into two stages --- unsupervised learning and supervised learning --- we achieve great flexibility in choosing learning algorithms. Using binarised-LapEig for the first stage and linear-SVM for the second stage, STH significantly outperforms binarised-LSI, LCH, and the state-of-the-art technique SpH \cite{Weiss2008}.
Since STH is a general learning framework, it is promising to achieve even higher effectiveness and efficiency if more powerful unsupervised or supervised learning algorithms can be employed.

We shall apply this technique to text mining tasks (such as automated text categorisation \cite{Sebastiani2002}) and content-based multimedia retrieval \cite{Lew2006} in the near future. 
It would also be interesting to combine semantic hashing and distributed computing (e.g., \cite{Lin2009}) to further improve the speed and scalability of similarity search.

\section*{Acknowledgements}
We are grateful to Dr Xi Chen (Alberta) for his valuable discussion and the London Mathematical Society (LMS) for their support of this work (SC7-09/10-6). 
We would also like to thank the anonymous reviewers for their helpful comments.

\bibliographystyle{abbrv}


\end{document}